\documentclass[pre,twocolumn,longbibliography]{revtex4-1}

\usepackage{amssymb,amsmath}

\usepackage{graphicx}

\newcommand\beq{\begin{equation}}
\newcommand\eeq{\end{equation}}
\newcommand\beqa{\begin{eqnarray}}
\newcommand\eeqa{\end{eqnarray}}

\newcommand{\al}{\alpha}

\begin{document}

\title{Mean square displacement of intruders in freely cooling multicomponent granular mixtures}

\author{Rub\'en G\'omez Gonz\'alez}
\email{Electronic address: ruben@unex.es}
\affiliation{Departamento de
Did\'actica de las Ciencias Experimentales y las Matem\'aticas, Universidad de Extremadura, E-10004 C\'aceres, Spain}

\author{Santos Bravo Yuste}
\email{Electronic address: santos@unex.es; \url{https://fisteor.cms.unex.es/investigadores/santos-bravo-yuste/}}
\author{Vicente Garz\'{o}}
\email{Electronic address: vicenteg@unex.es; \url{https://fisteor.cms.unex.es/investigadores/vicente-garzo-puertos/}}
\affiliation{Departamento de F\'{\i}sica and Instituto de Computaci\'on Cient\'{\i}fica Avanzada (ICCAEx), Universidad de Extremadura, E-06006 Badajoz, Spain}

\begin{abstract}
The mean square displacement (MSD) of intruders (tracer particles) immersed in a multicomponent granular mixture made up of smooth inelastic hard spheres in a homogeneous cooling state is explicitly computed. The multicomponent granular mixture consists of $s$ species with different masses, diameters, and coefficients of restitution. In the hydrodynamic regime, the time decay of the granular temperature of the mixture gives rise to a time decay of the intruder's diffusion coefficient $D_0$. The corresponding MSD of the intruder is determined by integrating the corresponding diffusion equation. As expected from previous works on binary mixtures, we find a logarithmic time dependence of the MSD which involves the coefficient $D_0$. To analyze the dependence of the MSD on the parameter space of the system, the diffusion coefficient is explicitly determined by considering the so-called second Sonine approximation (two terms in the Sonine polynomial expansion of the intruder's distribution function). The theoretical results for $D_0$ are compared with those obtained by numerically solving the Boltzmann equation by means of the direct simulation Monte Carlo method. We show that the second Sonine approximation improves the predictions of the first Sonine approximation, especially when the intruders are much lighter than the particles of the granular mixture.
In the long-time limit, our results for the MSD agree with those recently obtained by Bodrova [Phys. Rev. E \textbf{109}, 024903 (2024)] when $D_0$ is determined by considering the first Sonine approximation.


\end{abstract}

\date{\today}
\maketitle

\section{Introduction}
\label{sec0}


Granular materials in both nature and industry are usually characterized by some degree of polydispersity in both mass density and size. Due to the differences in mass and/or size of particles of each species, multicomponent mixtures exhibit
particle segregation or demixing \cite{OK00,K04}.
Thus, the study of transport properties in multicomponent granular mixtures is relevant not only from a fundamental point of view but also from a practical perspective.

When granular matter is subjected to an external excitation, the energy input supplied to the grains can compensate for the energy dissipated through collisions. In this situation, the motion of grains is quite similar to the chaotic motion of atoms or molecules in a conventional fluid. In this regime (referred to as rapid flow conditions), the collisions of grains play a relevant role in their dynamics and hence, kinetic theory tools (conveniently adapted to take into account the dissipative character of collisions) have been employed in the last few years \cite{G03,BP04,G19} to analyze granular flows. In particular, for sufficiently \emph{dilute} systems, granular materials can be modeled as a gas of smooth \emph{inelastic} hard spheres where the inelasticity of collisions is only characterized by a constant coefficient of normal restitution.

The fact that the collisions between grains are inelastic implies that their diffusive motion will eventually stop in the absence of an external excitation. For this reason, most of the studies of diffusion in granular systems have been carried out in driven steady states \cite{IALOB95,OU99,WHP01,MVPRKEU05,GASG23}. However, diffusion can also be studied in freely cooling systems, specifically in the homogeneous cooling state (HCS). The detailed analysis of diffusive transport in the HCS for multicomponent granular mixtures can be considered as an important goal
with some as yet open questions. For example, understanding how variations in particle sizes affect diffusion is crucial for optimizing pharmaceutical mixers.

To the best of our knowledge, the first works on the diffusion of an intruder in a freely cooling granular gas were focused on two limiting cases: (i) the self-diffusion problem (intruder is mechanically equivalent to the particles of the granular gas) \cite{BRCG00,BP00,BChChM15a} and the Brownian limit (intruder is much heavier than the particles of the granular gas) \cite{BRGD99,SVCP10}. In both limiting cases, Haff's law \cite{H83} for the granular temperature yields a diffusion coefficient that decays in time. This dependence gives rise to a logarithmic time dependence of the MSD of intruder. A more recent study has extended these previous findings to arbitrary values of the intruder-grain mass ratio \cite{ASG22}.

All previous works \cite{BRCG00,BP00,BChChM15a,BRGD99,SVCP10,ASG22} refer to the diffusion of the intruder in a granular gas. However, studies on intruder diffusion in multicomponent or polydisperse granular systems are much scarcer. We are aware of only a very recent work \cite{B24} where the MSD of granular particles in a multicomponent granular mixture (containing an arbitrary number of species with different masses and diameters) under HCS has been explicitly determined. Although this work fills an important gap in the granular literature, the results derived in Ref.\ \cite{B24} are based on a simplifying assumption that was not explicitly stated in the paper: the MSD of species $k$ was determined by considering the so-called first Sonine approximation to the partial diffusion coefficient $D_k$. However, as has been clearly shown in previous works on binary granular mixtures \cite{GM04,GV09,GV12,GMV13a,GG23,GG24}, the first Sonine approximation to $D_k$ exhibits significant  deviations from computer simulation data when the intruder is much lighter than the particles of the granular gas. These differences are clearly mitigated when considering the next correction to $D_k$, namely the second Sonine approximation to $D_k$. The question then arises as to whether and if so to what extent, the conclusions drawn from Ref.\ \cite{B24} may change when the second Sonine correction to $D_k$ is accounted for in the theory.

In this paper, we address the above question by determining the MSD of the intruder in a multicomponent granular mixture using both the first and second Sonine approximations for the diffusion coefficient. As occurs in binary systems under HCS \cite{GM04,GV09}, the comparison between
theory and the numerical results obtained from the direct simulation Monte Carlo (DSMC) method \cite{B94} shows that while the first and second approximations to the intruder diffusion coefficient $D_0$ yield practically the same results when the intruder is heavier than the particles of the multicomponent mixture, the second Sonine approximation provides a significant improvement over the first one when the intruder is lighter than the particles of the multicomponent mixture in the range of large inelasticity. It is also important to remark that our derivation of the MSD  is slightly different to the one provided in Ref.\ \cite{B24}. Nevertheless, we demonstrate that in the case that the intruder is mechanically equivalent to one of the species (let's say, for instance, the species $k$), our expression of the MSD of the species $k$ agrees with the expression obtained in Ref.\ \cite{B24} for a three-dimensional system when the diffusion coefficient $D_k$ is determined from the first Sonine approximation.

The plan of the paper is as follows. The analysis of the HCS of a multicomponent granular mixture composed of $s$ species is reviewed in Sec.\ \ref{sec1}.
As noted in previous works \cite{GD99b,MG02,BT02,DHGD02,BRM05,CHGG22}, the energy equipartition is broken down and the partial temperatures $T_i$ (which measure the mean kinetic energy of species $i$) differ from each other. On the other hand, at long times, the temperature ratios $T_i/T_j$ are independent of time.
The explicit form of the MSD of the intruders immersed in a multicomponent granular mixture is derived in Sec.\ \ref{sec2}. As expected \cite{ASG22}, the logarithmic time-dependence of the MSD arises from Haff's cooling law. Given that the MSD is written in terms of the intruder diffusion coefficient $D_0$, this transport coefficient is determined in Sec.\ \ref{sec3} by considering the first and second Sonine approximations. The theoretical predictions of $D_0$ from both approximations are compared against Monte Carlo simulations in Sec.\ \ref{sec4} in the case of a ternary mixture where one of the species is present in tracer concentration. The influence of the number of species in the multicomponent granular mixture on the tracer diffusion coefficient $D_0$ is analyzed in Sec.\ \ref{sec5} while the paper ends in Sec.\ \ref{sec6} with some concluding remarks.

\section{Multicomponent granular mixtures in HCS}
\label{sec1}

We consider an isolated multicomponent granular mixture of inelastic hard disks ($d=2$) or spheres ($d=3$) of masses $m_i$ and diameters $\sigma_i$ ($i=1,2,\ldots, s$). The subscript $i$ labels one of the $s$ mechanically different species or components and $d$ is the dimension of the system. For simplicity, we assume that the spheres are completely smooth; this means that the inelasticity of collisions between particles of species $i$ and $j$ is only characterized by the constant (positive) coefficient of normal restitution $\al_{ij}\leqslant 1$. The coefficient $\al_{ij}$ measures the ratio between the magnitude of the \emph{normal} component (along the line separating the centers of the two spheres at contact) of the relative velocity after and before the collision $i$-$j$. For relatively low densities, a kinetic theory description is appropriate, and the one-particle velocity distribution function $f_i(\mathbf{r}, \mathbf{v}, t)$ of species $i$ verifies the set of $s$-coupled nonlinear integro-differential Boltzmann kinetic equations \cite{G19}.

We assume that the granular mixture is in a spatially homogeneous state. In contrast to conventional (molecular) mixtures of hard spheres, there is no longer an evolution towards the Maxwellian distributions for $f_i$ since they are not a solution to the homogeneous set of Boltzmann equations. Instead, when one considers homogeneous initial conditions, there is a special solution which is reached after a few collision times: the so-called HCS solution \cite{NE98,GD99b}.    In the HCS, the set of Boltzmann equations read
\beq
\label{1.1}
\frac{\partial}{\partial t}f_i(\mathbf{v};t)=\sum_{j=1}^s J_{ij}[\mathbf{v}|f_i,f_j],
\eeq
where the Boltzmann--Enskog collision operator $J_{ij}$ is given by \cite{G19}
\beqa
\label{1.2}
& & J_{ij}[\mathbf{v}_1|f_i,f_j]=\sigma_{ij}^{d-1}\chi_{ij}\int d\mathbf{v}_2\int d\widehat{\boldsymbol{\sigma}}\Theta\left(\widehat{\boldsymbol{\sigma}}\cdot\mathbf{g}_{12}\right)\left(\widehat{\boldsymbol{\sigma}}
\cdot\mathbf{g}_{12}\right)\nonumber\\
& & \times \left[\al_{ij}^{-2}f_i(\mathbf{v}_1'';t)f_j(\mathbf{v}_2'';t)- f_i(\mathbf{v}_1;t)f_j(\mathbf{v}_2;t)\right].
\eeqa
Here, $\boldsymbol{\sigma}_{ij}=\sigma_{ij} \widehat{\boldsymbol{\sigma}}$, $\sigma_{ij}=(\sigma_i+\sigma_j)/2$, $\widehat{\boldsymbol{\sigma}}$ is a unit vector directed along the line of centers from the sphere of component $i$ to that of component $j$ at contact, $\Theta$ is the Heaviside step function, and $\mathbf{g}_{12}=\mathbf{v}_1-\mathbf{v}_2$ is the relative velocity of the colliding pair. Moreover, $\chi_{ij}$ refers to the pair correlation function for particles of species $i$ and $j$ when they are separated a distance $\sigma_{ij}$. In Eq.\ \eqref{1.2}, the relationship between the pre-collisional velocities $(\mathbf{v}_1'',\mathbf{v}_2'')$ and the post-collisional velocities $(\mathbf{v}_1,\mathbf{v}_2)$ is
\beq
\label{1.2.0}
\mathbf{v}_1''=\mathbf{v}_1-\mu_{ji}\left(1+\alpha_{ij}^{-1}\right)\left(\boldsymbol{\widehat{\sigma}}
\cdot\mathbf{g}_{12}\right)\boldsymbol{\widehat{\sigma}},
\eeq
\beq
\label{1.2.1bis}
\mathbf{v}_2''=\mathbf{v}_2+\mu_{ij}\left(1+\alpha_{ij}^{-1}\right)\left(\boldsymbol{\widehat{\sigma}}
\cdot\mathbf{g}_{12}\right)\boldsymbol{\widehat{\sigma}},
\eeq
where $\mu_{ij}=m_i/(m_i+m_j)$.
Note that in the HCS the Enskog equation becomes identical to the Boltzmann equation except for the presence of the pair correlation function $\chi_{ij}$.

The most relevant hydrodynamic field in the HCS is the granular temperature $T$. It is defined as
\beq
\label{1.2.1}
T=\frac{1}{d n}\sum_{i=1}^s\int d\mathbf{v}\; m_i v^2 f_i(\mathbf{v}),
\eeq
where $n=\sum_i n_i$ is the total number density and
\beq
\label{1.2.2}
n_i=\int d\mathbf{v}\; f_i(\mathbf{v})
\eeq
is the number density of species $i$. The balance equation of $T$ in the HCS is
\beq
\label{1.3}
\frac{\partial T}{\partial t}=-\zeta T,
\eeq
where the cooling rate $\zeta$ can be written as \cite{G19}
\beqa
\label{1.4}
\zeta(t)&=&\frac{\pi^{(d-1)/2}}{2d\Gamma\left(\frac{d+3}{2}\right)}\sum_{i=1}^s\sum_{j=1}^s\sigma_{ij}^{d-1}\chi_{ij} \frac{m_im_j}{m_i+m_j}\left(1-\alpha_{ij}^2\right)\nonumber\\
& & \times \int d\mathbf{v}_1\int d\mathbf{v}_2 {g}_{12}^3 f_i(\mathbf{v}_1;t)f_j(\mathbf{v}_2;t).
\eeqa
Although the relevant temperature at a hydrodynamic level is the global temperature $T$, it is also convenient to introduce the partial temperatures $T_i$ for each species. They measure the mean kinetic energy of species $i$. They are defined as
\beq
\label{1.4.1}
T_i=\frac{1}{d n_i}\int d\mathbf{v}\; m_i v^2 f_i(\mathbf{v}).
\eeq
According to Eqs.\ \eqref{1.2.1} and \eqref{1.4.1},
\beq
\label{1.4.2}
T=\sum_{i=1}^s\; x_i T_i,
\eeq
where $x_i=n_i/n$ is the mole fraction (or concentration) of species $i$.

For symmetry reasons, the mass and heat fluxes vanish in the HCS and the pressure tensor $P_{k\ell}=p\delta_{k\ell}$, where the hydrostatic pressure $p$ is \cite{GHD07}
\beq
\label{1.5}
p=n T \Big[1+\frac{\pi^{d/2}}{d\Gamma\left(\frac{d}{2}\right)}\sum_{i=1}^s\sum_{j=1}^s\; \mu_{ji}\; n\; \sigma_{ij}^d \chi_{ij} x_i x_j (1+\al_{ij})\gamma_i\Big],
\eeq
where $\gamma_i=T_i/T$ is the temperature ratio of species $i$. The rate of change of the partial temperatures $T_i(t)$ can be analyzed by the ``partial cooling rates'' $\zeta_i$. These quantities provide the rate of change of the mean kinetic energy of species $i$ due to collisions between themselves and with particles of different species ($j\neq i$). To obtain the evolution of $T_i$ one multiplies both sides of Eq.\ \eqref{1.1} by $\frac{m_i}{2}v^2$ and integrates over velocity. The result is
\beq
\label{1.6}
\frac{\partial T_i}{\partial t}=-\zeta_i T_i,
\eeq
where
\beq
\label{1.7}
\zeta_{i}(t)=\sum_{j=1}^s\zeta_{ij}(t), \quad \zeta_{ij}=-\frac{1}{d n_i T_i}\int d\mathbf{v} m_i v^2 J_{ij}[f_i,f_j].
\eeq
The total cooling rate $\zeta$ can be expressed in terms of the partial cooling rates $\zeta_i$ as
\beq
\label{1.8}
\zeta(t)=\sum_{i=1}^s\; x_i \gamma_i(t) \zeta_i(t).
\eeq
The time evolution of the temperature ratios $\gamma_i(t)$ can be easily derived from Eqs.\ \eqref{1.3} and \eqref{1.6} as
\beq
\label{1.9}
\frac{\partial \gamma_i}{\partial t}=\gamma_i\left(\zeta-\zeta_i\right), \quad i=1,\ldots,s.
\eeq

As said before, the term $\zeta_{ii}$ gives the contribution to the partial cooling rate $\zeta_i$ coming from the rate of energy loss from collisions between particles of the same species $i$. This term vanishes for elastic collisions but is different from zero when $\al_{ii}<1$. The remaining contributions $\zeta_{ij}$ ($i\neq j$) to $\zeta$ represent the transfer of energy between a particle of species $i$ and particles of species $j$. In general, $\zeta_{ij}\neq 0$  ($i\neq j$) for both elastic and inelastic collisions. However, for elastic collisions, when the distribution functions $f_i$ are Maxwellian distributions at the same temperature ($T_i=T$ for any species $i$), then $\zeta_{ij}=0$  ($i\neq j$). This occurs because of detailed balance, where the energy exchange between species is exactly countered by energy conservation for this state.

As widely discussed in Ref.\ \cite{GD99b}, the detailed balance state for inelastic collisions is the HCS. In this state, since the partial $\zeta_i$ and total $\zeta$ cooling rates never vanish, the partial $T_i$ and total $T$ temperatures are always time dependent. As for monocomponent granular gases \cite{NE98}, regardless of the initial uniform state considered, we expect that the Boltzmann--Enskog equation \eqref{1.1} tends towards the HCS solution where all the time dependence of the distributions $f_i(\mathbf{v};t)$  occurs only through the (total) temperature $T$. In this sense, the HCS solution qualifies as a \emph{normal} or hydrodynamic solution since the granular temperature $T(t)$ is in fact the relevant temperature at a hydrodynamic level. Thus, it follows from dimensional analysis that the distributions $f_i(\mathbf{v};t)$ have the form
\beq
\label{1.10}
f_i(\mathbf{v};t)=n_i v_\text{th}^{-d}(t)\varphi_i \left(\frac{\mathbf{v}}{v_\text{th}(t)}\right),
\eeq
where
\beq
\label{1.10.1}
v_\text{th}(t)=\sqrt{\frac{2T(t)}{\overline{m}}}
\eeq
is a thermal velocity defined in terms of the total temperature $T(t)$ of the mixture and $\overline{m}=\sum_{i=1}^s m_i/s$. In Eq.\ \eqref{1.10}, the temperature dependence of the reduced distributions $\varphi_i$ is through the dimensionless velocity $\mathbf{v}/v_\text{th}$. According to the definition \eqref{1.4.1} for the partial temperatures $T_i$ and the HCS solution \eqref{1.10} for $f_i$, it follows that all temperatures $\left\{T_1,\ldots, T_s, T\right\}$ are proportional to each other and their ratios are \emph{independent} of time. One possibility would be that $T_1=\ldots=T_s=T$, as happens in the case of molecular mixtures (elastic collisions). However, the ratios $\gamma_i$ ($i=1,\ldots,s$) must be determined by solving the set of Boltzmann--Enskog equations \eqref{1.1}. Results derived from kinetic theory \cite{GD99b}, computer simulations \cite{MG02,BT02,DHGD02,PMP02,BT02b,CH02,KT03,WJM03,BRM05,SUKSS06}, and even real experiments \cite{WP02,FM02} have clearly shown that the temperature ratios are in general different from 1; they exhibit in fact a complex dependence on the parameter space of the mixture.

Since the temperature ratios $\gamma_i$ achieve a time-independent value in the hydrodynamic regime, then according to Eq.\ \eqref{1.9}, the partial cooling rates $\zeta_i$ must be equal in the HCS:
\beq
\label{1.11}
\zeta_1(t)=\zeta_2(t)=\cdots =\zeta_s(t)=\zeta(t).
\eeq
The last identity in Eq.\ \eqref{1.11} is based on the fact that $\zeta=\zeta_i\sum_i x_i \gamma_i=\zeta_i$.
The constraint \eqref{1.11} allows us to determine the $s-1$ independent temperature ratios $\gamma_i$.

The left hand side of the Boltzmann--Enskog equation  \eqref{1.1} can be more explicitly written when one takes into account Eq.\  \eqref{1.10} for the distributions $f_i$:
\beq
\label{1.12}
\frac{\partial f_i}{\partial t}=\frac{\partial f_i}{\partial T}\frac{\partial T}{\partial t}=\frac{1}{2}\zeta\frac{\partial}{\partial \mathbf{v}}\cdot \left(\mathbf{v}f_i\right).
\eeq
Therefore, in dimensionless form, Eq.\ \eqref{1.9} reads
\beq
\label{1.13}
\frac{1}{2}\zeta_i^* \frac{\partial}{\partial \mathbf{c}}\cdot \left(\mathbf{c}\varphi_i\right)=\sum_{j=1}^s\; J_{ij}^*[\mathbf{c}|\varphi_i,\varphi_j],
\eeq
where
\beq
\label{1.13.1}
\zeta_i^*=\frac{\zeta_i}{\nu}, \quad \mathbf{c}=\frac{\mathbf{v}}{v_\text{th}}, \quad J_{ij}^*[\varphi_i,\varphi_j]=\frac{v_\text{th}^d}{n_i \nu}J_{ij}[f_i,f_j].
\eeq
 Here,
\beq
\label{1.14}
\nu(t)=n\overline{\sigma}^{d-1}v_\text{th}(t)
\eeq
is an effective collision frequency and $\overline{\sigma}=\frac{1}{s}\sum_{i=1}^s\sigma_i$. The use of $\zeta_i^*$ instead of $\zeta^*=\zeta/\nu$ on the left hand side of Eq.\ \eqref{1.13} is allowed by Eq.\  \eqref{1.11}; this choice is more convenient since the first few velocity moments of Eq.\ \eqref{1.13} are directly obtained without any specification of the distributions $\varphi_i$.

Therefore, we are in front of a well-possed mathematical problem since we have to solve the set of $s$ Boltzmann--Enskog equations \eqref{1.1} for velocity distribution functions $f_i(v;t)$ of the form \eqref{1.10} and subject to the $s-1$ constraints \eqref{1.11}. These $2s-1$ equations must be solved to determine the $s$ distributions $f_i$ and the $s-1$ temperature ratios $\gamma_i$. As in the case of monocomponent granular gases, approximate expressions for the above quantities are obtained by considering the first few terms of the expansion of the distributions $f_i$ in a series of Sonine (or Laguerre) polynomials \cite{GD99b,CHGG22}.

\subsection{Haff 's cooling law}

According to Eq.\ \eqref{1.7}, explicitly computing the cooling rates $\zeta(t)$ requires to know the velocity distributions $f_i(\mathbf{v};t)$. However, since in the HCS the time dependence of $\zeta$ is solely through $\sqrt{T}$, the cooling rate can be written as $\zeta(t)=\nu(t) \zeta^*$, where $\zeta^*=\zeta_i^*=\sum_j\zeta_{ij}^*$ is a time-independent quantity [see for instance, the Maxwellian approximation to $\zeta_{ij}^*$ given by Eq.\ \eqref{1.17}]. Thus, the integration of Eq.\ \eqref{1.3} can be easily carried out and the result is
\beq
\label{1.15}
T(t)=\frac{T(0)}{\left(1+\frac{1}{2}\zeta(0)t\right)^2},
\eeq
where $T(0)$ is the initial temperature and $\zeta(0)$ denotes the cooling rate at $t=0$. Equation \eqref{1.15} is known as Haff's cooling law for the HCS \cite{H83}. Its form is formally identical to the one derived for a monocomponent granular gas, except that $\zeta(0)$ refers to the initial total cooling rate. According to Eq.\ \eqref{1.11}, $\zeta_1(0)=\cdots=\zeta_s(0)=\zeta(0)$.

\subsection{Maxwellian approximation to $\varphi_i$}

So far, all the results are exact. However, to obtain the explicit dependence of the temperature ratios on the parameter space of the system, one needs to know the scaled distributions $\varphi_i(\mathbf{c})$. For binary \cite{MG02,BT02,DHGD02,BRM05} and ternary \cite{CHGG22} mixtures, it has been shown that the temperature ratios $\gamma_i$ can be well estimated by replacing $\varphi_i(\mathbf{c})$ by its Maxwellian form, i.e.,
\beq
\label{1.16}
\varphi_i(\mathbf{c})\to \pi^{-d/2} \theta_i^{d/2}
e^{-\theta_i c^2}, \quad \theta_i=\frac{m_i T}{\overline{m}T_i}.
\eeq
In the Maxwellian approximation \eqref{1.16}, the dimensionless quantities $\zeta_{ij}^*=\zeta_{ij}/\nu$ are given by \cite{G19}
\beqa
\label{1.17}
& & \zeta_{ij}^*=\frac{4\pi^{(d-1)/2}}{d\Gamma\left(\frac{d}{2}\right)}x_j \chi_{ij}\left(\frac{\sigma_{ij}}{\overline{\sigma}}\right)
^{d-1}\mu_{ji}(1+\al_{ij})\nonumber\\
& & \times \left(\frac{\theta_i+\theta_j}{\theta_i\theta_j}\right)^{1/2}
\Big[1-\frac{1}{2}\mu_{ji}(1+\al_{ij})\frac{\theta_i+\theta_j}{\theta_i\theta_j}\Big].
\nonumber\\
\eeqa
The (dimensionless) partial cooling rates $\zeta_i^*$ can be easily obtained from Eqs.\ \eqref{1.7} and \eqref{1.17}. Substitution of the forms of $\zeta_i^*$ into the identities \eqref{1.11} yields in general nonlinear algebraic equations for $\gamma_i$ whose numerical solutions provide the dependence of the temperature ratios on the parameter space of the system.

\section{Mean square displacement of intruders in a multicomponent granular mixture}
\label{sec2}

As usual in the study of the MSD, let us assume that some impurities or intruders of mass $m_0$ and diameter $\sigma_0$ are added to the multicomponent granular mixture. The intruders are in general mechanically different from any of the species of the mixture. Let us denote by $\al_{0i}$ the coefficient of restitution for collisions between the intruder and particles of the species $i$. In general, $\al_{0i}\neq \al_{ij}$ for any species $i$ and $j$.

Since the concentration of intruders is negligible, the state of the mixture constituted by $s$ species is not perturbed and hence the HCS is still preserved. Formally, the resulting system can be seen as a granular mixture of $s+1$ species where one of the species is present in tracer concentration.
For the sake of conciseness,  we will thereafter refer to this system as consisting of an intruder immersed in a multicomponent granular mixture.

Under these conditions, the velocity distribution function $f_0(\mathbf{r}, \mathbf{v}; t)$ of the intruders obeys the kinetic equation
\beq
\label{2.0}
\frac{\partial f_0}{\partial t}+\mathbf{v}\cdot \nabla f_0=\sum_{i=1}^s\; J_{0i}[f_0,f_i],
\eeq
where the Enskog--Lorentz collision operator $J_{0i}[f_0,f_i]$ gives the rate of change of $f_0$ due to the \emph{inelastic} collisions between the intruders and particles of the species $i$. It is given by
\beqa
\label{2.0.1}
& & J_{0i}[\mathbf{v}_1|f_0,f_i]=\sigma_{0i}^{d-1}\chi_{0i}\int d\mathbf{v}_{2}\int d\widehat{\boldsymbol {\sigma
}}\Theta (\widehat{{\boldsymbol {\sigma}}} \cdot {\mathbf g}_{12}) (\widehat{\boldsymbol {\sigma }}\cdot {\mathbf g}_{12})\nonumber\\
& & \times
\left[\al_{0i}^{-2}f_0({\mathbf v}_{1}'')f_i({\mathbf v}_{2}'')-f_0({\mathbf v}_{1})f_i({\mathbf v}_{2})\right].
\eeqa
As in Eq.\ \eqref{1.2}, $\mathbf{g}_{12}=\mathbf{v}_1-\mathbf{v}_2$ is the relative velocity, $\widehat{\boldsymbol {\sigma }}$ is a unit vector, and $\Theta$ is the Heaviside step function. In addition, $\chi_{0i}$ is the pair correlation function for intruders and particles of the species $i$, and $\sigma_{0i}=(\sigma_0+\sigma_i)/2$. In Eq.\ \eqref{2.0.1}, the relationship between $(\mathbf{v}_1'', \mathbf{v}_2'')$ and $(\mathbf{v}_1, \mathbf{v}_2)$ is
\beq
\label{2.0.2}
\mathbf{v}_1''=\mathbf{v}_1-(1+\al_{0i}^{-1})\mu_{i0}(\widehat{{\boldsymbol {\sigma }}} \cdot {\mathbf g}_{12})\widehat{\boldsymbol {\sigma }},
\eeq
\beq
\label{2.0.3}
\mathbf{v}_2''=\mathbf{v}_2+(1+\al_{0i}^{-1})\mu_{0i}(\widehat{{\boldsymbol {\sigma }}} \cdot {\mathbf g}_{12})\widehat{\boldsymbol {\sigma}},
\eeq
where
\beq
\label{2.0.3.1}
\mu_{i0}=\frac{m_i}{m_i+m_0}, \quad
\mu_{0i}=\frac{m_0}{m_i+m_0}.
\eeq
In a similar way, the collision rules for the direct collision $(\mathbf{v}_1, \mathbf{v}_2)\to (\mathbf{v}_1', \mathbf{v}_2')$ with the same collision vector $\widehat{{\boldsymbol {\sigma }}}$ are defined as
\beq
\label{2.0.4}
\mathbf{v}_1'=\mathbf{v}_1-(1+\al_{0i})\mu_{i0}(\widehat{{\boldsymbol {\sigma }}} \cdot {\mathbf g}_{12})\widehat{\boldsymbol {\sigma }},
\eeq
\beq
\label{2.0.5}
\mathbf{v}_2'=\mathbf{v}_2+(1+\al_{0i})\mu_{0i}(\widehat{{\boldsymbol {\sigma }}} \cdot {\mathbf g}_{12})\widehat{\boldsymbol {\sigma}}.
\eeq

Since the intruder may freely lose or gain momentum and energy in its interactions with the multicomponent mixture, these quantities are not invariants of the (inelastic) Enskog--Lorentz collision operator $J_{0i}[f_0,f_i]$. Only the number density of intruders
\beq
\label{2.1}
n_0(\mathbf{r};t)=\int d\mathbf{v} f_0(\mathbf{r}, \mathbf{v};t)
\eeq
is conserved. The continuity equation for $n_0$ can be easily obtained from the kinetic equation \eqref{2.0} as
\beq
\label{2.2}
\frac{\partial n_0}{\partial t}=-\nabla \cdot \mathbf{j}_0,
\eeq
where
\beq
\label{2.3}
\mathbf{j}_0(\mathbf{r}; t)= \int d\mathbf{v}\; \mathbf{v}\; f_0(\mathbf{r}, \mathbf{v};t)
\eeq
is the intruder particle flux.

Equation \eqref{2.2} becomes a closed differential equation for the intruder number density when one expresses the flux $\mathbf{j}_0$ in terms of $n_0$. As usual, a constitutive equation for $\mathbf{j}_0$ can be obtained by solving the Enskog--Lorentz kinetic equation \eqref{2.0} by means of the Chapman--Enskog method \cite{CC70} conveniently adapted to dissipative dynamics. To first order in $\nabla n_0$, the constitutive equation for $\mathbf{j}_0$ is
\beq
\label{2.4}
\mathbf{j}_0=-D_0 \nabla n_0,
\eeq
where $D_0(t)$ is the time-dependent intruder diffusion coefficient. Substitution of Eq.\ \eqref{2.4} into Eq.\ \eqref{2.3} yields the diffusion equation
\beq
\label{2.5}
\frac{\partial x_0}{\partial t}=D_0(t) \nabla^2 x_0,
\eeq
where $x_0=n_0/n$
is the concentration of the intruder particles. As we know, in contrast to the usual diffusion equation for molecular (elastic) gases, Eq.\ \eqref{2.5} cannot be directly integrated in time because of the time dependence of the diffusion coefficient $D_0$. However, for times much longer than the mean free time (hydrodynamic regime), $D_0(t)$ adopts a form where its dependence on time is only through its dependence on the granular temperature $T(t)$ \cite{BRCG00,GM04,G19}. In addition, kinetic theory calculations \cite{GM04} show that $D_0(t)$ can be expressed as follows:
\beq
\label{2.6}
D_0(t)=\frac{T(t)}{m_0 \nu(t)} D_0^*,
\eeq
where the (dimensionless) diffusion transport coefficient $D_0^*$ depends on the parameter space of the system but it is a time-independent quantity. An explicit, albeit approximate, expression for the reduced diffusion coefficient $D_0^*$ can be obtained by considering for instance the first and second Sonine approximations to the Chapman--Enskog solution. The explicit form of $D_0^*$ will be provided in Sec.\ \ref{sec3}.

As usual in freely cooling mixtures \cite{ASG22}, the time dependence of the diffusion equation $D_0(t)$ can be eliminated by introducing a set of appropriate dimensionless time $\tau$ and space $\mathbf{r}'$ variables:
\beq
\label{2.10}
\tau=\int_0^t dt' \nu(t'), \quad \mathbf{r}'=\frac{\mathbf{r}}{\ell}.
\eeq
Here,
\beq
\label{2.11}
\ell=\frac{1}{n\overline{\sigma}^{d-1}}
\eeq
is a unit length (proportional to the mean free path of a monocomponent molecular gas of hard spheres) and the dimensionless time variable $\tau$ measures the effective (average) number of collisions per gas particle in the time interval between 0 and $t$.
An explicit formula for $\tau(t)$ is readily obtained by making use of Haff's law \eqref{1.15} in the expression  \eqref{1.14} of $\nu(t)$ in terms of the thermal velocity $v_{th}(t)$. The time integral defining $\tau(t)$ then gives
\beq
\label{2.12}
\tau(t)=2\frac{\nu(0)}{\zeta(0)} \ln \left(1+\frac{1}{2}\,\frac{\zeta(0)}{\nu(0)} t^*\right),
\eeq
where $t^*=\nu(0) t$. Note that for a multicomponente granular mixture
\beq
\label{2.13}
\frac{\nu(0)}{\zeta(0)}=\frac{1}{\zeta^*}=\frac{1}{\zeta_i^*}=\frac{1}{\sum_{j=1}^s\zeta_{ij}^*},
\eeq
where $\zeta_{ij}^*$ is given by Eq.\ \eqref{1.17} in the Maxwellian approximation.

In terms of the variables $\tau$ and $\mathbf{r}'$, the diffusion equation \eqref{2.5} becomes
\beq
\label{2.14}
\frac{\partial x_0}{\partial \tau}=\widetilde{D}_0\nabla_{\mathbf{r}'}^2 x_0,
\eeq
where $\nabla_{\mathbf{r}'}^2$ is the Laplace operator in the $\mathbf{r}'$ coordinate and $\widetilde{D}_0$ is the dimensionless diffusion coefficient
\beq
\label{2.15}
\widetilde{D}_0=
\frac{1}{2}\frac{\overline{m}}{m_0}D_0^*.
\eeq
As expected, Eq.\ \eqref{2.15} is thus a standard diffusion equation with a \emph{time-independent} diffusion coefficient $\widetilde{D}_0$. It follows that the MSD of the \emph{intruder's position} $r'$ after a \emph{time} interval $\tau$ is
\beq
\label{2.16}
\langle |\Delta \mathbf{r}'|^2(\tau)\rangle=2 d \widetilde{D}_0 \tau,
\eeq
with $\Delta \mathbf{r}'\equiv \mathbf{r}'(\tau)-\mathbf{r}'(0)$.  Then,
\beq
\label{2.17}
\frac{\partial}{\partial \tau}\langle |\Delta \mathbf{r}'|^2(\tau)\rangle=2 d \widetilde{D}_0.
\eeq
In terms of the original variables  $\mathbf{r}$ and $t$, one has
\beq
\label{2.18}
\frac{\partial}{\partial t}\langle |\Delta \mathbf{r}|^2(t)\rangle=2 d \frac{T(t)}{m_0\nu(t)} D_0^*.
\eeq
Equation \eqref{2.18} can be seen as a generalization of the Einstein formula relating the diffusion coefficient to the MSD. In terms of the unit length $\ell$, the MSD can be written as
\beq
\label{2.19}
\langle |\Delta \mathbf{r}|^2(t)\rangle=2d \frac{\overline{m}}{m_0}\frac{D_0^*}{\zeta^*}\ln \left(1+ \frac{1}{2}\zeta^*t^*\right) \ell^2.
\eeq
Under the assumptions made (hydrodynamic solution restricted to first-order in $\nabla n_0$), Eq.\ \eqref{2.19} is exact and very general, but $D_0^*$ and $\zeta^*$ need to be explicitly determined. In the case of $\zeta^*$, as mentioned before, a good estimate of it  is provided by the Maxwellian approximation \eqref{1.17}. In the case of $D_0^*$, we will compute it by considering the two first terms in a Sonine polynomial expansion of the first-order Chapman--Enskog solution to the  distribution function $f_0(\mathbf{r}, \mathbf{v};t)$.

When the intruder and particles of the multicomponent mixture are  mechanically equivalent (i.e, when $m_0=m_i$, $\sigma_0=\sigma_i$, and $\alpha_{0i}=\alpha_{ij}$, $i,j=1,\cdots,s$), Eq.\ \eqref{2.19} agrees with previous results derived for the self-diffusion problem \cite{BChChM15,BChChM15a}. Equation \eqref{2.19} also extends to multicomponent mixtures the results obtained in Ref.\ \cite{ASG22} for binary systems.

It is quite apparent that for inelastic collisions Eq.\ \eqref{2.19} shows that the MSD increases logarithmically with time. This means that the diffusion of the intruder is ultraslow, namely, it is even slower than in the case of subdiffusion. As occurs for binary systems \cite{ASG22}, the time-dependent argument of the logarithm of Eq.\ \eqref{2.19} is independent of the mechanical properties of the intruder. This is essentially due to the fact that the time dependence of the MSD is directly obtained from the Haff's cooling law \eqref{1.15}, which only depends on the properties of the multicomponent granular mixture through the initial total cooling rate $\zeta(0)$.

\section{Determination of the diffusion coefficient $D_0$}
\label{sec3}

The goal of this section is to determine the diffusion coefficient $D_0$ by means of the Chapman--Enskog method \cite{CC70}. The analysis follows similar steps as those previously made in the case of a binary mixture \cite{GM04}. Thus, in the first-order in $\nabla n_0$, the first-order distribution function $f_0^{(1)}(\mathbf{r}, \mathbf{v};t)$ in the Chapman--Enskog solution is given by
\beq
\label{3.1}
f_0^{(1)}(\mathbf{v})=\boldsymbol{\mathcal{A}}_0(\mathbf{v})\cdot \nabla n_0,
\eeq
where $\boldsymbol{\mathcal{A}}_0(\mathbf{v})$ verifies the integral equation
\beq
\label{3.2}
-\zeta T\partial_T \boldsymbol{\mathcal{A}}_0-\sum_{i=1}^s J_{0i}\left[\boldsymbol{\mathcal{A}}_0,f_i\right]=-\frac{f_0^{(0)}}{n_0}\mathbf{v},
\eeq
and the expression of the collision operator $J_{0i}\left[\boldsymbol{\mathcal{A}}_0,f_i\right]$ can be easily inferred from Eq.\ \eqref{2.0.1}.

The diffusion coefficient $D_0$ is defined as
\beq
\label{3.4}
D_0=-\frac{1}{d}\int d\mathbf{v}\; \mathbf{v}\cdot \boldsymbol{\mathcal{A}}_0(\mathbf{v}).
\eeq
As for elastic collisions \cite{CC70}, the linear integral equation \eqref{3.2} may be approximately solved by expanding the unknown $\boldsymbol{\mathcal{A}}_0(\mathbf{v})$ in a Sonine poynomial expansion. Here, we determine $D_0$ by considering contributions to $\boldsymbol{\mathcal{A}}_0$ up to the second Sonine approximation. In this approach, $\boldsymbol{\mathcal{A}}_0$ reads
\beq
\label{3.5}
\boldsymbol{\mathcal A}_0(\mathbf{v})\to -f_{0,\text{M}}(\mathbf{v})\Big[a_1\mathbf{v}+a_2 \mathbf{S}_0(\mathbf{v})\Big],
\eeq
where $f_{0,\text{M}}(\mathbf{v})$ is the Maxwellian distribution function
\beq
\label{3.6}
f_{0,\text{M}}(\mathbf{v})=n_0\left(\frac{m_0}{2\pi T_0}\right)^{d/2}\exp\left(-\frac{m_0 v^2}{2 T_0}\right),
\eeq
and $\mathbf{S}_0(\mathbf{v})$ is the polynomial
\beq
\label{3.7}
\mathbf{S}_0(\mathbf{v})=\Big(\frac{1}{2}m_0 v^2-\frac{d+2}{2}T_0\Big)\mathbf{v}.
\eeq
The Sonine coefficients $a_1$ and $a_2$ are defined as
\beq
\label{3.8}
a_1=-\frac{m_0}{d n_0 T_0}\int d\mathbf{v}\; \mathbf{v}\cdot \boldsymbol{\mathcal A}_0(\mathbf{v})=\frac{m_0 D_0}{n_0 T_0},
\eeq
\beq
\label{3.9}
a_2=-\frac{2}{d(d+2)}\frac{m_0}{n_0 T_0^3}\int d\mathbf{v}\; \mathbf{S}_0(\mathbf{v})\cdot \boldsymbol{\mathcal A}_0(\mathbf{v}).
\eeq
The evaluation of the coefficients $a_1$ and $a_2$ is carried out in the Appendix \ref{appA}.

The expression of the reduced diffusion coefficient $D_0^*=(m_0\nu/T)D_0$ depends on the Sonine approximation considered. The first Sonine approximation $D_0^*[1]$ to $D_0^*$ is
\beq
\label{3.10}
D_0^*[1]=\frac{\gamma_0}{\nu_{a,0}^*-\frac{1}{2}\zeta^*},
\eeq
where $\gamma_0=T_0/T$ and the expression of the (dimensionless) collision frequency $\nu_{a,0}^*$ is displayed in the Appendix \ref{appB}. The second Sonine approximation $D_0^*[2]$ to $D_0^*$ is given by
\beq
\label{3.11}
D_0^*[2]=\frac{\gamma_0\left(\nu_{d,0}^*-\frac{3}{2}\zeta^*\right)}{(\nu_{a,0}^*-\frac{1}{2}\zeta^*)(\nu_{d,0}^*-
\frac{3}{2}\zeta^*)-\nu_{b,0}^*(\nu_{c,0}^*-\zeta^*)}.
\eeq
Here, the expressions of the (reduced) collision frequencies $\nu_{a,0}^*$, $\nu_{b,0}^*$, $\nu_{c,0}^*$, and $\nu_{d,0}^*$ are also provided in the Appendix \ref{appB}.

\subsection{Comparison with the results derived in Ref.\ \cite{B24}}

We assume that the intruder is mechanically equivalent to one of the species; let's denote by $k$ this species. In this case, $D_0^*=D_k^*$ and according to Eq.\ \eqref{2.19} the MSD of species $k$ is
\beqa
\label{3.12}
\langle R_k^2(t) \rangle&\equiv&
\langle |\Delta\mathbf{r}|^2(t)\rangle_k
\nonumber\\
&=&2d \frac{\overline{m}}{m_k}\frac{D_
k^*}{\zeta^*}\ln \left(1+ \frac{1}{2}\zeta^*t^*\right) \ell^2.
\eeqa
To compare with the MSD derived in Ref.\ \cite{B24}, let us write Eq.\ \eqref{3.12} in terms of the initial diffusion coefficient $D_k(0)$:
\beq
\label{3.13}
D_k(0)=\frac{T(0)}{m_k \nu(0)}D_k^*,
\eeq
where use has been of the fact that in the HCS the (dimensionless) diffusion coefficient $D_k^*$ is independent of time.
The explicit form of $D_k^*$ depends on the Sonine approximation considered. Taking into account that
\beq
\label{3.14}
\ell^2=\left(\frac{v_\text{th}(0)}{\nu(0)}\right)^2=\frac{2T(0)/\overline{m}}{\nu(0)^2},
\eeq
the MSD of species $k$ can be rewritten as
\beq
\label{3.15}
\langle R_k^2(t) \rangle=2 d \tau_0 D_k(0) \ln \left(1+\frac{t}{t_0}\right),
\eeq
where $t_0=2/\zeta^* \nu(0)$. The first Sonine approximation $D_k^*[1]$ to $D_k^*$ is
\beq
\label{3.16}
D_k^*[1]=\frac{\gamma_k}{\nu_{a,k}^*-\frac{1}{2}\zeta^*},
\eeq
where we recall that $\zeta^*=\zeta_i^*=\sum_{j=1}^s \zeta_{ij}^*$ is given by Eq.\ \eqref{1.17} and
\beqa
\label{3.17}
\nu_{a,k}^*&=&\frac{4\pi^{(d-1)/2}}{d\Gamma\left(\frac{d}{2}\right)}\sum_{i=1}^s x_i \chi_{ki}\left(\frac{\sigma_{ki}}{\overline{\sigma}}\right)^{d-1}
\mu_{ik}\nonumber\\
& & \times
\Bigg(\frac{\theta_i+\theta_k}{\theta_i\theta_k}\Bigg)^{1/2}\frac{1+\al_{ik}}{2}.
\eeqa
In the three-dimensional case ($d=3$), Eq.\ \eqref{3.15}  agrees with Eq.\ (64) of Ref.\ \cite{B24} (which provides the MSD at long times)
 when $D_k^*$ is evaluated by the first Sonine approximation, Eq.\ \eqref{3.16}\footnote{Note that the evaluation of the
scaling  time $t_0$ involves quantities associated with \emph{all} the species of the multicomponent mixture since $\zeta^*=\zeta_i^*=\sum_{j=1}^s \zeta_{ij}^*$. However, in Ref.\ \cite{B24}, the quantity $t_0$ is defined in Eq.\ (28),  which is restricted to a monocomponent granular gas.}.

\section{Comparison between theory and Monte Carlo simulations}
\label{sec4}

It is quite apparent that the theoretical results displayed along Sec.\ \ref{sec3} for the diffusion coefficient $D_0$ are approximate since they have been obtained by considering one (first Sonine approximation) or two (second Sonine approximation) terms in the Sonine polynomial expansion of $\boldsymbol{\mathcal{A}}_0(\mathbf{V})$. Reliability of both approaches can be assessed via a comparison with computer simulations. Here, as in previous works for binary systems \cite{GM04,GV09}, we numerically solve the Boltzmann equation for dilute granular mixtures ($n \sigma^d\to 0$ and so, $\chi_{ij}\to 1$) by means of the DSMC method \cite{B94}. Given that the diffusion of intruders in a granular gas (binary mixture where one of the species is present in tracer concentration) has been widely analyzed in the above papers \cite{GM04,GV09}, we consider here a ternary system, namely, the diffusion of intruders in a granular binary mixture ($s=2$).

The adaptation of the DSMC method to the case of granular mixtures has been described in previous works (see, for instance, Ref.\ \cite{MG02}). We will only mention in this section the aspects related to our specific problem: diffusion of intruders (or tracer particles) in a granular binary mixture under HCS. Thus, in the tracer limit  ($x_0=n_0/(n_1+n_2)\to 0$) of a ternary mixture constituted by species $0$, $1$, and $2$, collisions $0-0$ are not considered. In addition, when collisions $0-1$ and $0-2$ take place, the post-collisional velocities from the scattering rule [given by Eqs.\ \eqref{2.0.4} and \eqref{2.0.5}] are only assigned to the intruder particle $0$. In this context, the number of particles of each species $N_i$ has only a statistical meaning and so, they can be chosen arbitrarily.

Two different stages are distinguished during the simulations. In the first stage, starting from a certain initial state, the system (intruders and particles of the granular binary mixture) evolves towards the HCS state. Once the system has reached the HCS state (second stage), the kinetic temperatures $T_i(t)$ and the diffusion coefficient $D_0(t)$ are measured. As usual, the coefficient $D_0(t)$ is obtained from the MSD of intruders [Eq.\ \eqref{2.18}], i.e.,
\beq
\label{4.1}
D_0(t)=\frac{1}{6\Delta t}\Big[\langle |\mathbf{r}_i(t+\Delta t)-\mathbf{r}_i(0)|^2\rangle-
\langle |\mathbf{r}_i(t)-\mathbf{r}_i(0)|^2\rangle\Big],
\eeq
where a three-dimensional ($d=3$) system has been considered. In Eq.\ \eqref{4.1}, $|\mathbf{r}_i(t)-\mathbf{r}_i(0)|$ is the distance traveled by the intruder from $t=0$ until time $t$; $t=0$ being the beginning of the second stage. In addition, the average $\langle \ldots \rangle$ is done over the $N_0$ intruders and $\Delta t$ is the time step.

\begin{figure}[t]
{\includegraphics[width=0.85\columnwidth]{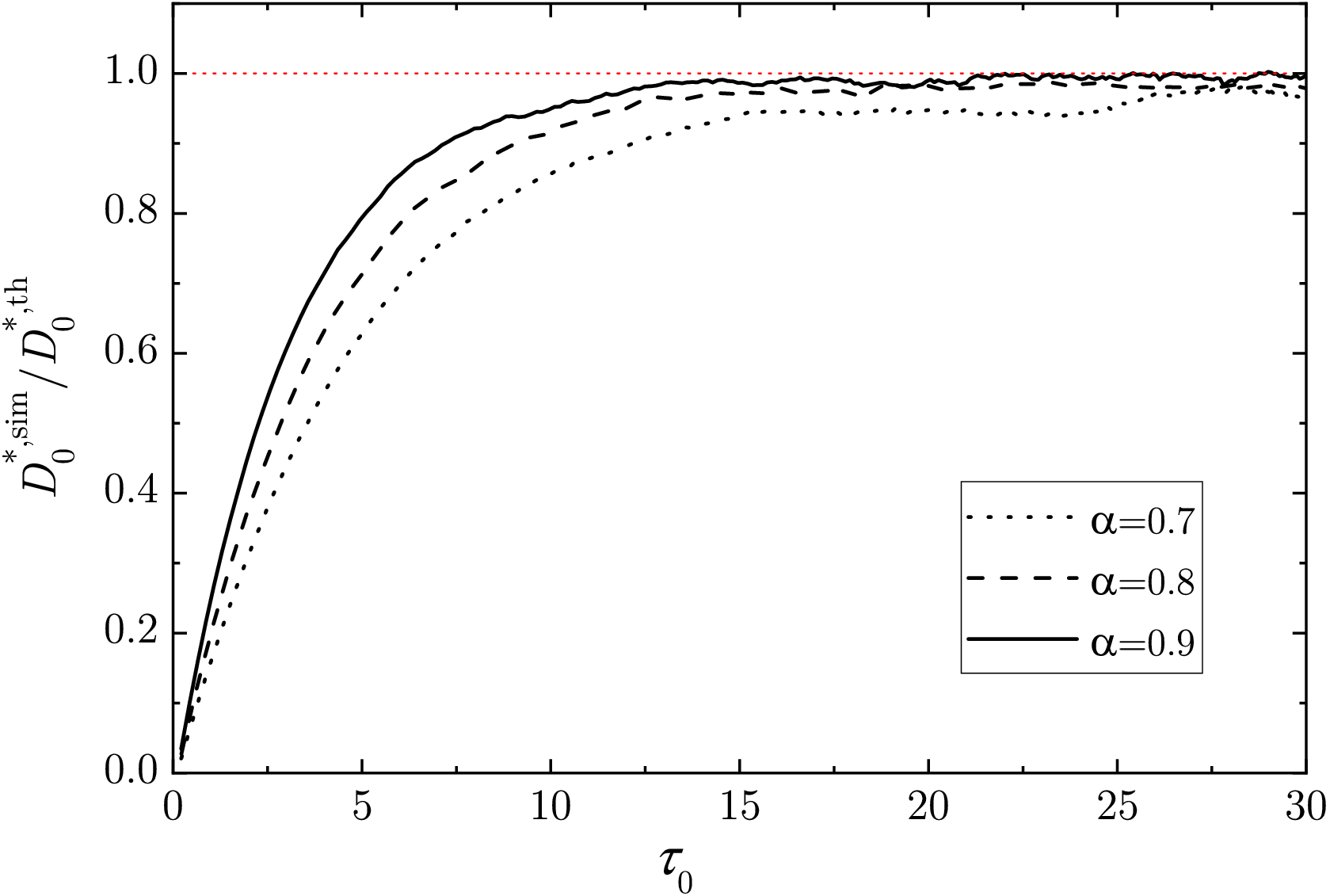}}
\caption{Dependence of the ratio $D_0^\text{*,sim}/D_0^\text{*,th}$ on the (dimensionless) time $\tau_0$ for a ternary mixture with $x_1=0.5$, $m_0/m_1=3$, $m_0/m_2=2$,  $\sigma_0=\sigma_1=\sigma_2$, and three different values of the (common) coefficient of restitution $al_{ij}\equiv \al$: 0.9, 0.8, and 0.7. Here, $D_0^\text{*,sim}$ refers to the simulation result of the dimensionless tracer diffusion coefficient $D_0^*$ while $D_0^\text{*,th}$ corresponds to its theoretical result obtained from the first Sonine approximation.}
\label{fig1}
\end{figure}

According to Eq.\ \eqref{2.6}, the validity of a hydrodynamic description implies necessarily that the time dependence of the diffusion coefficient $D_0(t)$ only occurs via the square root of the granular temperature $T(t)$ ($D_0(t)\propto \sqrt{T(t)}$ for hard spheres). Thus, the dimensionless diffusion coefficient $D_0^*$ must reach a constant value independent of time after a transient regime. Our simulation results clearly show that $D_0^*$ reaches a stationary value in all the systems simulated in the work. To illustrate this behaviour, we plot in Fig.\ \ref{fig1} the ratio $D_0^\text{*,sim}/D_0^\text{*,th}$ versus the (dimensionless) time $\tau_0$, defined as the average number of collisions experienced by the tracer particles up to a time $t$. The relationship between $\tau_0(t)$ and $\tau(t)$ is given in the Appendix \ref{appC}. In Fig.\ \ref{fig1}, we consider a ternary mixture with $x_1=n_1/(n_1+n_2)=0.5$, $m_0/m_1=3$, $m_0/m_2=2$,  $\sigma_0=\sigma_1=\sigma_2$, and three different values of the (common) coefficient of restitution $\al_{ij}\equiv \al$: 0.9, 0.8, and 0.7. Moreover, $D_0^\text{*,sim}$ refers to the simulation result of the dimensionless tracer diffusion coefficient $D_0^*$ while $D_0^\text{*,th}$ corresponds to its theoretical result obtained from the first Sonine approximation. As occurs for binary systems (see Fig.\ 3 of Ref.\ \cite{GM04}), we observe that after a certain number of collisions, $D_0^\text{*,sim}/D_0^\text{*,th}$ reaches a time-independent plateau, which value fluctuates around 1. This means that the value of the coefficient $D_0^*$ measured in the simulations agrees very well with the one obtained from the Boltzmann kinetic equation. Additionally, we can observe that the transition to the steady state takes more time the lower the value of $\alpha$. This is because, in these specific simulations, the initial values of the particle velocities are sampled from Maxwellian distributions at the same temperature. As the inelasticity increases, the breakdown of energy equipartition becomes more pronounced, and therefore, the longer it takes for the velocity distributions to relax to their stationary forms.

\begin{table}[t]
\caption{Three-dimensional systems considered in the computer simulations}
\label{table1}
\begin{center}
\begin{tabular}{|c|c|c|c|c|c|}
\hline
Case&$x_1$&$\frac{\sigma_0}{\sigma_1}$&$\frac{\sigma_0}{\sigma_2}$&$\frac{m_0}{m_1}$&$\frac{m_0}{m_2}$\\ \hline
I&0.5&1&1&3&2\\ \hline
II&0.8&1&1&$\frac{1}{2}$&$\frac{1}{4}$ \\ \hline
III&$0.5$&$2^{1/3}$&$5^{1/3}$&2&5 \\ \hline
IV&$0.9$&$\frac{10}{3}$&$\frac{1}{2}$&$(\frac{10}{3})^{3}$&$\frac{1}{8}$ \\ \hline
\end{tabular}
\end{center}
\end{table}

\begin{figure}[t]
{\includegraphics[width=0.85\columnwidth]{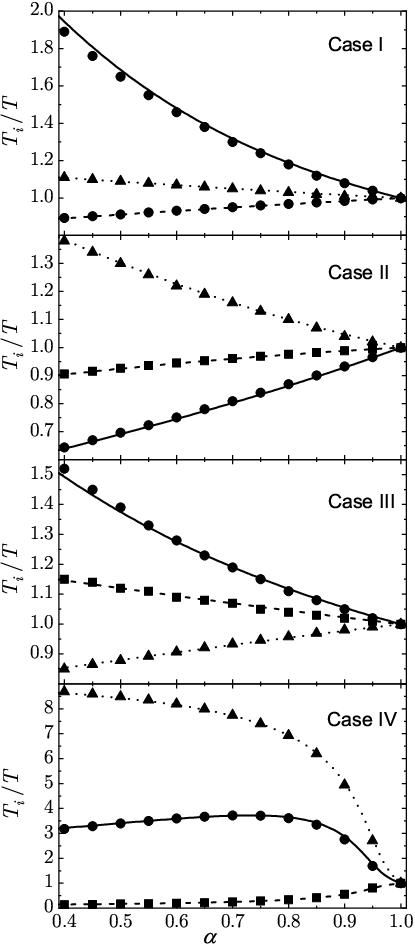}}
\caption{Temperature ratios $T_i/T$ versus the (common) coefficient of restitution $\al$ for the cases I, II, III, and IV described in Table \ref{table1}. The solid lines and circles correspond to $T_0/T$, the dashed lines and squares refer to $T_1/T$, while the dotted lines and triangles correspond to $T_2/T$. Symbols are the DSMC results while lines are the theoretical results. }
\label{fig2}
\end{figure}

\begin{figure}[t]
{\includegraphics[width=0.85\columnwidth]{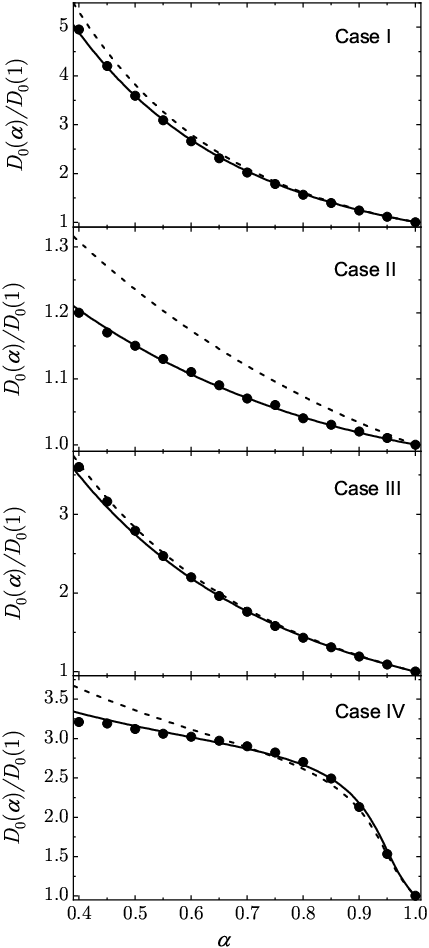}}
\caption{Plot of the (reduced) tracer diffusion coefficient $D_0(\al)/D_0(1)$ as a function of the (common) coefficient of restitution $\al$ for the cases I, II, III, and IV described in Table \ref{table1}. The solid and dashed lines refer to the theoretical results obtained from the second and first Sonine approximations, respectively, while the symbols (circles) refer to the DSMC results. Here, $D_0(1)$ denotes the elastic-limit value of the coefficient $D_0$ consistently obtained in each approximation.
}
\label{fig3}
\end{figure}

Four different cases or systems have been considered in the simulations. In all of them we consider a three-dimensional system with a common coefficient of restitution $\al_{ij}\equiv \al$. The values of concentration, masses, and diameters of the different systems are displayed in Table \ref{table1}. In the two first cases (Cases I and II), we have considered a ternary system with identical diameters ($\sigma_0=\sigma_1=\sigma_2$) but different masses. In the cases III and IV, we have assumed that intruders and particles of the species 1 and 2 have the same mass density, namely, $m_0/m_1=(\sigma_0/\sigma_1)^3$ and $m_0/m_2=(\sigma_0/\sigma_2)^3$. Furthermore, since we are essentially interested in studying the influence of dissipation on the tracer diffusion coefficient, we have normalized $D_0(\al)$ with respect to its value for elastic collisions ($\al_{ij}=1$). Thus, the steady values of the ratio $D_0(\al)/D_0(1)$ obtained from simulation data are compared against their theoretical predictions obtained by considering the first (Eq.\ \eqref{3.10}) and second Sonine (Eq.\ \eqref{3.11}) approximations. The theoretical elastic value $D_0(1)$ has been consistently obtained in each approximation.

Before considering the diffusion coefficient, we study first the dependence of the temperature ratios $\gamma_0= T_0/T$, $\gamma_1= T_1/T$, and $\gamma_2= T_2/T$ on the coefficient of restitution. Note that in the tracer limit $\gamma_2=(1-x_1 \gamma_1)/(1-x_1)$, so that only two of the temperature ratios are independent.
Figure \ref{fig2} shows the dependence of the temperature ratios on inelasticity for the cases considered in Table \ref{table1}. The ratios $\gamma_i$ are obtained as described at the end of Sec.\ II.B.
As expected, is it quite apparent that the total energy is not equally distributed between the different species. This means that there is a breakdown of energy equipartition in granular mixtures. As mentioned in Sec.\ \ref{sec1}, the lack of energy equipartition in granular mixtures has been confirmed in many computer simulation works \cite{MG02,DHGD02,BT02,BT02b,CH02,KT03,WJM03,BRM05,BRM06,SUKSS06,BR11,BLB14,VLSG17,BOB20} as well as in real experiments of agitated \cite{WP02,FM02} and freely cooling \cite{PTHS24} mixtures. In addition, the lack of energy equipartition has dramatic consequences in the case of a large impurity/gas mass ratio since there is a peculiar ``phase transition'' with one phase where the diffusion coefficient grows without bound \cite{SD01,SD01a}.

As occurs for binary systems, it is quite apparent in Fig.\ \ref{fig2} that for large differences in the mass ratios $m_0/m_1$ and $m_0/m_2$ the temperature ratios depart significantly from 1, even for relatively weak dissipation (say for instance, $\al \gtrsim 0.8$). We also observe that in general the temperature of the heavier species is larger than that of the lighter ones. Moreover, in spite that the temperature ratios $\gamma_i$ have been estimated by considering the simple Maxwellian approximations \eqref{1.16} to the scaled distributions $\varphi_i$, the theoretical predictions for $\gamma_i$ exhibit in general an excellent agreement with the DSMC results, even for strong inelasticities.

We consider now the (dimensionless) diffusion coefficient $D_0(\al)/D_0(1)$. Based on the previous results obtained for a binary system \cite{GM04}, one would expect that the first and second Sonine approximations practically coincide in the Rayleigh gas limit (namely, when the mass and/or the diameter of the intruder is larger than that of the granular gas particles) while both approximations appreciably differ in the Lorentz gas limit (namely, when the mass and/or the diameter of the intruder is smaller than that of the granular gas particles).
Figure \ref{fig3} shows the $\al$-dependence of $D_0(\al)/D_0(1)$ for the four cases displayed in Table \ref{table1}. In case I, the mass of the intruder is larger than the mass of the particles of the granular gas mixture (a Rayleigh-like scenario).  It is quite apparent from Fig.\ \ref{fig3} that both Sonine approximations (Sonine of order 1 and Sonine of order 2) are close to each other in this case, although the second approximation is clearly better than the first one, especially when the collisional inelasticity is large (let's say for instance, $\al \gtrsim 0.6$).
Case II is the opposite of case I: now the intruder mass is smaller than that of the particles of the granular mixture (a Lorentz-like scenario). We observe that the two Sonine approximations differ significantly, especially for strong inelasticity. In addition, we see that while the first Sonine approximation clearly overestimates the computer simulation data, the agreement of the second Sonine approximation with the results of the DSMC is excellent in the complete range of values of $\al$ analyzed.
In case III, the mass density of all particles (intruder and gas grains) is the same, but the particles of the granular mixture are larger and heavier than the intruder.  This case is thus closer to the Rayleigh limit, and again we find that the two Sonine approximations give quite similar results.
Finally, in case IV, the mass density of all particles is again the same, but now the grains of species 1 are larger and heavier than the intruder, while the grains of species 2 are smaller and lighter than the intruder. However, since most of the grains are of species 1 (since $x_1=0.9$), the scenario is more Rayleigh-like than Lorenz-like. As expected, we find that the two Sonine approximations give similar results, although again the second Sonine solution is the best, providing again a  very good agreement with the DSMC results.



In summary, our results for the diffusion of intruders in a binary granular mixture (ternary system) share similarities with those previously reported \cite{GM04} for a binary system:  The second Sonine approximation clearly improves on the first Sonine approximation, especially when the intruders  are lighter than the particles of the surrounding granular binary mixture. In the latter case, the first Sonine solution for $D_0$ clearly overestimates the simulation data, while the second Sonine approximation agrees very well with the DSMC results  even for large values of the inelasticity.

\section{Tracer diffusion in a multicomponent mixture}
\label{sec5}

Although the theoretical expressions \eqref{3.10} and \eqref{3.11} of the first and second Sonine approximations, respectively, for the tracer diffusion coefficient hold for an arbitrary number of species, the results presented in Sec.\ \ref{sec4} have focused on the case of a ternary mixture (intruders immersed in a granular binary mixture). In the present section, we analyze the effect of increasing the polydispersity (number of species in the mixture) on the diffusion coefficient $D_0$ for three-dimensional systems.

It is obvious that a complete study of the dependence of $D_0$ on the parameter space of the system is quite complex, mainly due to the large number of parameters involved in the description of highly polydisperse granular systems. Here, as in Figs.\ \ref{fig2} and \ref{fig3}, we consider a three-dimensional system with a common coefficient of restitution $\al_{ij}\equiv \al$. In addition, since we are mainly interested in evaluating the effect of sizes and masses of the polydisperse system on diffusion, we assume equimolar granular mixtures ($x_1=x_2=\cdots=x_s=1/s$). Only the second Sonine approximation \eqref{3.11} is considered throughout this section to analyze the dependence of the (reduced) tracer diffusion coefficient $D_0(\al)/D_0(1)$ on the diameter and mass ratios, and the coefficient of restitution.

\begin{figure}[t]
{\includegraphics[width=0.85\columnwidth]{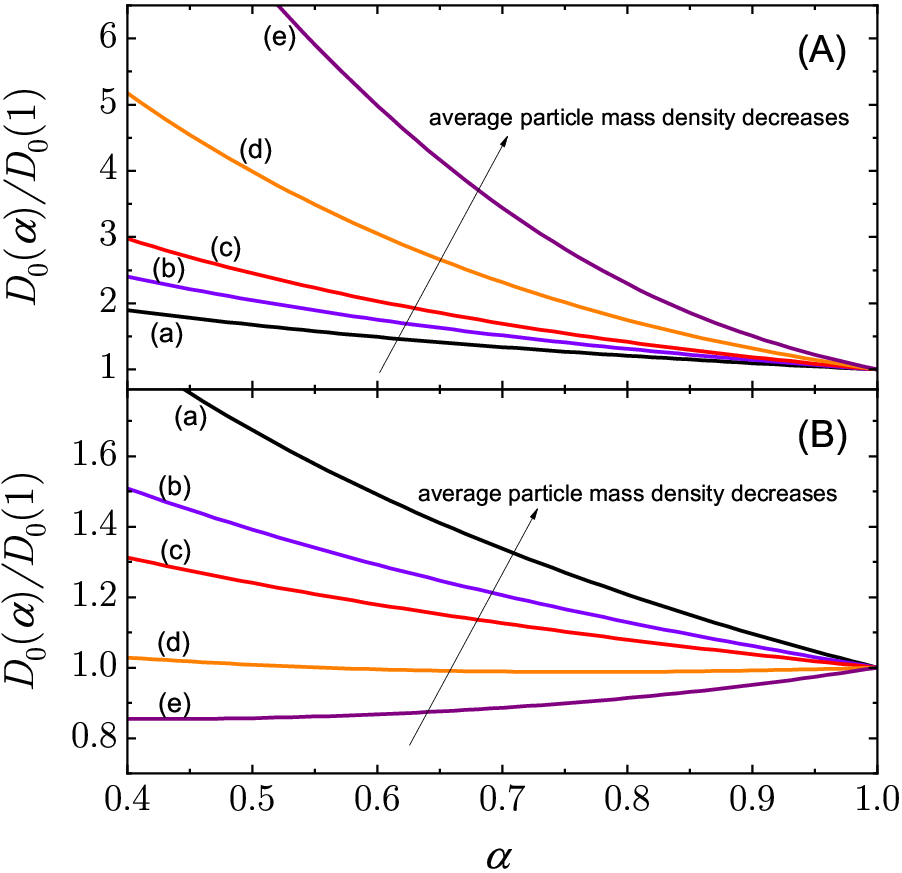}}
\caption{Plot of the (reduced) tracer diffusion coefficient $D_0(\al)/D_0(1)$ as a function of the (common) coefficient of restitution $\al$ for  equimolar mixtures with $s$ species of the same size, $\sigma_0/\sigma_i=1$. The solid lines are obtained from the second Sonine approximation \eqref{3.11} to $D_0$.
In panel (A) $m_0/m_i=i$  and in panel (B)  $m_0/m_i=1/i$ with $i=1,2,\ldots,s$. In both panels  (a) $s=1$, (b) $s=2$, (c) $s=3$, (d) $s=6$, (e) $s=10$.
}
\label{fig4}
\end{figure}

\begin{figure}[th]
{\includegraphics[width=0.85\columnwidth]{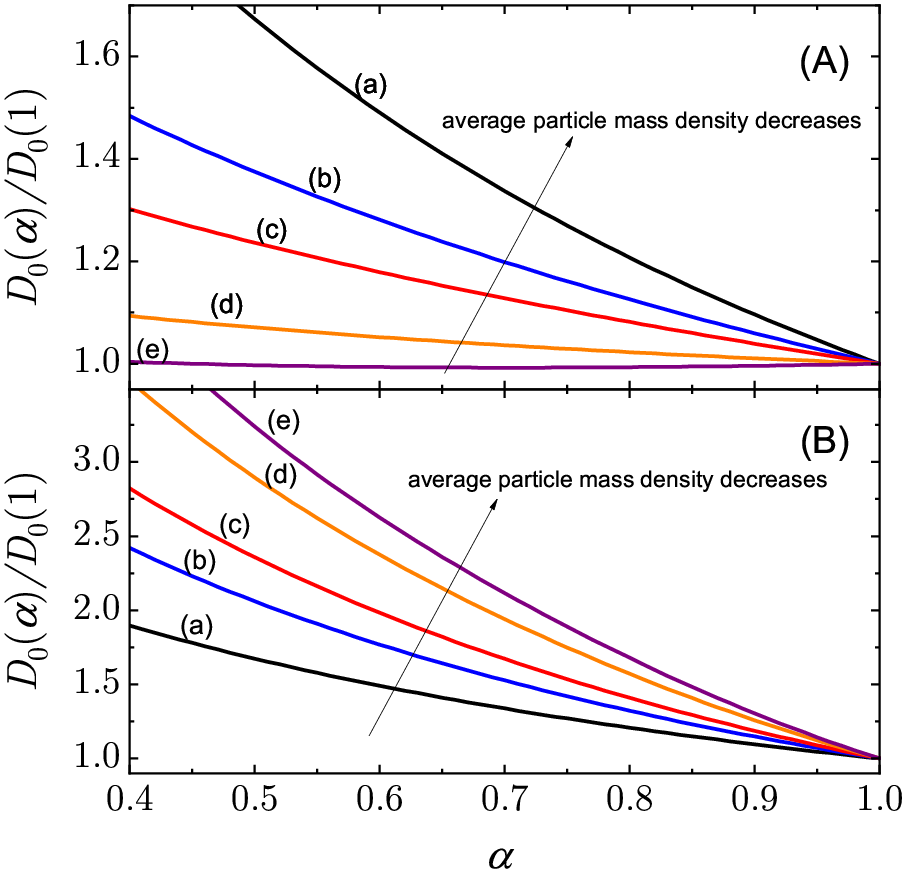}}
\caption{Plot of the (reduced) tracer diffusion coefficient $D_0(\al)/D_0(1)$ as a function of the (common) coefficient of restitution $\al$ for mixtures of $s$ species with the same mass, $m_0/m_i=1$. The solid lines are obtained from the second Sonine approximation \eqref{3.11} to $D_0$. In panel (A) $\sigma_0/\sigma_i=i$  and in panel (B)  $\sigma_0/\sigma_i=1/i$ with $i=1,2,\ldots,s$. In both panels  (a) $s=1$, (b) $s=2$, (c) $s=3$, (d) $s=6$, (e) $s=10$.
}
\label{fig5}
\end{figure}

Our results are shown in Figs.\ \ref{fig4} and \ref{fig5}.
In Fig.~ \ref{fig4}, we fix the diameter ratio of the mixture ($\sigma_0/\sigma_i=1$) and consider different values of the mass ratios $m_0/m_i$ $(i=1,2,\cdots,s$). In Fig.~\ref{fig5}, we fix the mass ratio ($m_0/m_i=1$) and consider different values of the diameter ratios $\sigma_0/\sigma_i$ $(i=1,2,\cdots,s$). The combination of both plots allows us to measure the effect of the diameter and mass ratios on the diffusion of intruders in a granular mixture of $s$ species.

Panels A and B of Fig.\ \ref{fig4} show the dependence of the (reduced) tracer diffusion coefficient $D_0(\al)/D_0(1)$ on the restitution coefficient $\al$ for $\sigma_0/\sigma_i=1$ and mixtures consisting of different numbers of species ($s=1,2,3,6,$ and 10) with different values of the mass ratio. In panel A of Fig.\ \ref{fig4}, the mass distribution  is  $m_0/m_i=i$ with $i=1,2,\ldots,s$.
In panel B of Fig.\ \ref{fig4}, the mass distribution is  $m_0/m_i=1/i$ with $i=1,2,\ldots,s$.

We observe in Fig.\ \ref{fig4} that
the deviations of the diffusion coefficient $D_0(\al)$ from its form for elastic collisions increase with increasing dissipation when the intruder is heavier than the particles of the multicomponent mixture. However, according to panel B, the above tendency is not maintained in the case of highly polydisperse granular mixtures  when the intruder is lighter than the particles of the mixture (see the case $s=10$). In this case,  the ratio $D_0(\al)/D_0(1)$ has a non-monotonic dependence on $\al$. Furthermore, for a given value of the coefficient of restitution $\alpha$, it is quite obvious that, when we add particles lighter than the intruders to the mixture,  the (scaled) diffusion coefficient $D_0(\al)/D_0(1)$ increases, while the opposite happens when the particles are heavier than the intruders.

In panels A and B of Fig.\ \ref{fig5}, we illustrate the diffusion of the intruders in a granular mixture where the mass ratio $m_0/m_i=1$ and the diameter distribution is defined as $\sigma_0/\sigma_i=i$ (panel A) and $\sigma_0/\sigma_i=1/i$ (panel B) with $i=1,2,\ldots,s$. As in Fig.\ \ref{fig4}, we have considered $s=1,2,3,6,$ and 10. We observe different trends from those found in Fig.\ \ref{fig4}. Thus, for a given value of $\al$, $D_0(\al)/D_0(1)$ decreases (increases) when we add particles to the mixture that are smaller (larger) than the intruders. This means that intruder's diffusion is enhanced (hindered) with respect to its elastic value by adding particles to the mixture that are smaller (larger) than the intruders.

For a fixed value of the restitution coefficient, the results shown in Figs.\ \ref{fig4} and \ref{fig5} can be easily summarized in terms of the average particle mass density $\overline{\rho}_p$ of the mixture: $\overline{\rho}_p= \sum_{i=1}^s \rho_{i,p}/s$ with $\rho_{i,p}=m_i/(\pi \sigma_i^3/6)$.  We see that the lighter the mixture (i.e. the smaller $\overline{\rho}_p$), the larger $D_0(\al)/D_0(1)$.
On the other hand, for a given mixture, \emph{in general}, $D_0(\al)/D_0(1)$ increases as inelasticity increases (i.e., as $\alpha$ decreases).
 However, this does not hold as more and more species with increasing mass density $\rho_i$ are added to the mixture.  In this case we find that $D_0(\al)/D_0(1)$ eventually decreases as $\alpha$ decreases: see line (e) of panel B of Fig.\ \ref{fig4} (this also happens with line (e) of panel A of Fig.\ \ref{fig5}, although it is hardly noticeable due to the scale).

\section{Concluding remarks}
\label{sec6}

The determination of the MSD of an intruder in a freely cooling granular mixture is a very interesting and not completely understood problem. There are likely two different reasons for which the problem is quite complex. First, there is a large number of relevant parameters involved in the description of diffusion in granular mixtures. Second, there is also a wide array of complexities that arise during the derivation of kinetic theory models. Thus, to gain some insight into the general problem, the first studies consider two limiting cases: (i) when the intruder is mechanically equivalent to the particles of the granular gas (self-diffusion problem) \cite{BRCG00,BP00,BChChM15a}, and (ii) when the intruder is much more heavier than the particles of the granular gas (Brownian limiting case) \cite{BRGD99,SVCP10}. One of the main conclusions in these studies is that the MSD of an intruder presents a logarithmic time dependence. According to Haff's law \cite{H83}, the origin of this logarithmic  dependence arises from the algebraic decay of the granular temperature. These studies have been extended more recently to arbitrary values of the intruder-grain mass and/or diameter ratios \cite{ASG19} and even when the system is immersed in a molecular gas (granular suspension) \cite{GASG23}.

Very recently \cite{B24}, Bodrova has extended all the previous attempts to the case of granular mixtures with an arbitrary number of mechanically different species. On the other hand, to determine the expression of the MSD of the species $k$, one needs to know the diffusion coefficient $D_k(t)$. This coefficient is given in terms of the unknown $\boldsymbol{\mathcal{A}}_k(\mathbf{v})$ which is in fact the solution of the linear integral equation \eqref{3.2}. Thus, as occurs for molecular hard spheres mixtures \cite{CC70}, the above integral equation cannot be exactly solved (except for the so-called inelastic Maxwell models \cite{BK03}) and hence, one has to expand $\boldsymbol{\mathcal{A}}_k(\mathbf{v})$ in a Sonine polynomial expansion. In Ref.\ \cite{B24}, although not explicitly stated in the paper, the simplest first Sonine approximation (a first order polynomial in the particle velocity $\mathbf{v}$) to $\boldsymbol{\mathcal{A}}_k(\mathbf{v})$ is only considered. Although this approach can be accurate in some cases (for binary systems, when the intruder is much heavier than the particles of the granular gas), it yields a significant disagreement with computer simulations when the intruder is much lighter than the granular gas particles. In contrast, as has been widely shown in several papers for binary systems \cite{GM04,GV09,GV12,GMV13a,GG23,GG24}, the second Sonine approximation to the diffusion coefficient  improves the predictions of the first Sonine approximation since it leads in most of the cases to an excellent agreement with computer simulations. The determination of the tracer diffusion coefficient in a multicomponent granular mixture from the first and second Sonine approximations has been one of the main goals of the present contribution.

To assess the reliability of the Sonine approximations, we have also numerically solved the Boltzmann equation by means of the DSMC method \cite{B94} conveniently adapted to dissipative dynamics. Our simulations have been focused on the case of a ternary mixture where one of the species is present in tracer concentration. Four different sort of systems (displayed in Table \ref{table1}) have been considered.
As happens for binary systems \cite{GM04}, the comparison between theory and computer simulations clearly shows the superiority of the second Sonine solution over the first one,
especially in case II  (mass of intruders smaller than that of the particles of the granular gas) for strong inelasticity.
In addition,  the excellent agreement found between the second Sonine approximation to $D_0$ and DSMC results (see Fig.\ \ref{fig3}) confirms the accuracy of this theoretical expression for conditions of practical interest in mixtures constituted by three species. We expect that this good agreement will be kept in mixtures containing more than three species.

We want also to remark that the derivation carried out here for the MSD is slightly different to the one provided in Ref.\ \cite{B24}. While in the above work \cite{B24} the MSD is obtained from the velocity correlation function, here we identify the MSD of an intruder immersed in a granular mixture through the diffusion equation. In any case, when the intruder is mechanically equivalent to one of the species (let's say species $k$), then the MSD of the species $k$ derived here agrees with the one obtained in Ref.\ \cite{B24} when the first Sonine approximation to $D_k$ is considered.

Although we have focused our study on the case of ternary systems, our theory also applies to highly polydisperse granular systems. This allows us to examine the effect on intruder diffusion within mixtures when additional species of different masses and sizes are introduced. Specifically, we considered equimolar mixtures containing 1, 2, 3, 6, and 10 mechanically distinct species. As clearly shown in Figs.\ \ref{fig4} and \ref{fig5}, the influence of varying the (average) particle mass density of the mixture on the intruder diffusion coefficient $D_0$  is generally quite significant. When particles and intruders share a common coefficient of restitution, $\alpha$, the diffusion coefficient $D_0$ relative to its value for elastic collisions ($\alpha=1$) increases as the (average) particle mass density of the mixture decreases. For a given mixture, the diffusion coefficient typically also increases as $\alpha$ decreases; however, this trend does not hold when the average particle mass density of the mixture is sufficiently high.

One of the main limitations of the present work is its restriction to \emph{smooth} inelastic hard spheres. This means that inelaticity in collisions only affects to the translational degrees of freedom. The extension of the results derived here to the case of inelastic \emph{rough} hard spheres is an interesting open problem. Although some attempts have been made in the self-diffusion problem \cite{BB12}, it still remains to assess the impact of particles' roughness on diffusion when the intruder and particles of the granular gas are mechanically different. Another possible project may be the extension of the present results to granular suspensions where the influence of the interstitial fluid on grains can be modeled via a drag force plus a stochastic-like Langevin term \cite{GASG23,BW24}. Work along these lines is in progress.

\acknowledgments

We acknowledge financial support from Grant No. PID2020-112936GB-I00 funded by MCIN/AEI/ 10.13039/501100011033.

\appendix

\section{First and second Sonine approximations to the diffusion coefficient $D_0$}
\label{appA}

In this appendix we give some technical details on the determination of the Sonine coefficients $a_1$ and $a_2$. Substitution of
Eq.\ \eqref{3.5} into the integral equation \eqref{3.2} yields
\begin{widetext}
\beq
\label{a1}
\zeta T\partial_T \Big(a_1 f_{0\text{M}}\mathbf{v}+a_2 f_{0\text{M}}\mathbf{S}_0\Big)
+a_1\sum_{i=1}^sJ_{0i}[f_{0\text{M}}\mathbf{v},f_i]+a_2\sum_{i=1}^sJ_{0i}[f_{0\text{M}}\mathbf{S}_0,f_i]=-\frac{f_0^{(0)}}{n_0} \mathbf{v}.
\eeq
\end{widetext}
Next, we multiply Eq.\ \eqref{a1} by $\mathbf{v}$ and integrate over the velocity. The result is
\beq
\label{a2}
\left(\nu_{a,0}-\frac{1}{2}\zeta\right)D_0+\frac{n_0 T_0^2}{m_0} \nu_{b,0}\, a_2=\frac{T_0}{m_0},
\eeq
where use has been made of the identity $a_1=(m_0 D/n_0 T_0)$ and we have taken into account that $a_1 T \propto T^{1/2}$ and so $T\partial_T a_1 T=(1/2)a_1 T$. Moreover, we have introduced the quantities
\beq
\label{a3}
\nu_{a,0}=\sum_{i=1}^s \nu_{a,0i}, \quad \nu_{a,0i}=-\frac{m_0}{d n_0 T_0}\int d\mathbf{v}\; \mathbf{v}\cdot J_{0i}[f_{0\text{M}}\mathbf{v},f_i],
\eeq
\beq
\label{a4}
\nu_{b,0}=\sum_{i=1}^s \nu_{b,0i}, \quad \nu_{b,0i}=-\frac{m_0}{d n_0 T_0^2}\int d\mathbf{v}\; \mathbf{v}\cdot J_0[f_{0\text{M}}\mathbf{S}_0,f_i].
\eeq
If only the first Sonine correction is retained ($a_2=0$), the solution to Eq.\ \eqref{a2} is simply
\beq
\label{a5}
D_0[1]=\frac{T_0/m_0}{\nu_{a,0}-\frac{1}{2}\zeta}.
\eeq
Equation \eqref{a5} leads to Eq.\ \eqref{3.10}.

To close the problem, one has to multiply Eq.\ \eqref{a2} by $\mathbf{S}_0(\mathbf{v})$ and integrate over $\mathbf{v}$. After some algebra, one achieves the result
\beq
\label{a6}
\Big(\nu_{c,0}-\zeta\Big)\frac{m_0 D_0}{n_0 T_0^2}+\left(\nu_{d,0}-\frac{3}{2}\zeta\right)a_2=0,
\eeq
where
\beq
\label{a6.0}
\nu_{c,0}=\sum_{i=1}^s \nu_{c,0i}, \quad \nu_{d,0}=\sum_{i=1}^s \nu_{d,0i}.
\eeq
Here,
\beq
\label{a7}
\nu_{c,0i}=-\frac{2}{d(d+2)}\frac{m_0}{n_0 T_0^2}\int d\mathbf{v}\; \mathbf{S}_0\cdot J_{0i}[f_{0\text{M}}\mathbf{v},f_i],
\eeq
\beq
\label{a8}
\nu_{d,0i}=-\frac{2}{d(d+2)}\frac{m_0}{n_0 T_0^3}\int d\mathbf{v}\; \mathbf{S}_0\cdot J_{0i}[f_{0\text{M}}\mathbf{S}_0,f_i].
\eeq

In reduced units and using matrix notation, Eqs.\ \eqref{a2} and \eqref{a6} can be rewritten as
\beq
\label{a9}
\left(
\begin{array}{cc}
\nu_{a,0}^*-\frac{1}{2}\zeta^*&\gamma_0^2 \nu_{b,0}^*\\
\frac{\nu_{c,0}^*-\zeta^*}{\gamma_0^2}&\nu_{d,0}^*-\frac{1}{2}\zeta^*
\end{array}
\right)
\left(
\begin{array}{c}
D_0^*\\
a_2^*
\end{array}
\right)
=
\left(
\begin{array}{c}
\gamma_0\\
0
\end{array}
\right).
\eeq
Here, $D_0^*=(m_0 \nu/T)D_0$, $a_2^*=n_0 T \nu a_2$, $\nu_{a,0}^*=\nu_{a,0}/\nu$, $\nu_{b,0}^*=\nu_{b,0}/\nu$, $\nu_{c,0}^*=\nu_{c,0}/\nu$, and $\nu_{d,0}^*=\nu_{d,0}/\nu$.
The solution to Eq.\ \eqref{a9} gives the expression \eqref{3.11} of the second Sonine approximation $D_0^*[2]$ to $D_0^*$.

\section{Reduced collision frequencies}
\label{appB}

To obtain the explicit dependence of $D_0^*[2]$ and $D_0^*[1]$ on the parameter space of the system, one still needs to determine the quantities $\nu_{a,0}^*$, $\nu_{b,0}^*$, $\nu_{c,0}^*$, and $\nu_{d,0}^*$. According to Eqs.\ \eqref{a3}, \eqref{a4}, \eqref{a6.0}, \eqref{a7}, and \eqref{a8}, the above reduced collision frequencies are given in terms of the quantities
$\nu_{a,0i}^*$, $\nu_{b,0i}^*$, $\nu_{c,0i}^*$, and $\nu_{d,0i}^*$. These quantities have been evaluated in previous works when the distribution $f_i$ is approximated by the Maxwellian distribution given by Eq.\ \eqref{c3}.

To display the final expressions, it is convenient to introduce the quantities
\beq
\label{b2}
\theta_{0}=\frac{m_0 T}{\overline{m} T_0},  \quad
\lambda_{0i}=\mu_{0i}\theta_i-\mu_{i0}\theta_0,
\eeq
and $\theta_i$ defined in Eq.\ \eqref{1.16}. In terms of these quantities,   $\nu_{a,0i}^*$, $\nu_{b,0i}^*$, $\nu_{c,0i}^*$, and $\nu_{d,0i}^*$ are given, respectively, by
\begin{widetext}
\begin{equation}
\label{a12}
\nu_{a,0i}^*=\frac{2\pi^{(d-1)/2}}{d\Gamma\left(\frac{d}{2}\right)}x_i\left(\frac{\sigma_{0i}}{\overline{\sigma}}
\right)^{d-1}\chi_{0i}\mu_{i0} (1+\alpha_{0i}) \left(\frac{\theta_0+\theta_{i}}{\theta_{i}\theta_0}\right)^{1/2},
\end{equation}
\begin{equation}
\label{a13}
\nu_{b,0i}^*=\frac{\pi^{(d-1)/2}}{d\Gamma\left(\frac{d}{2}\right)}x_i\left(\frac{\sigma_{0i}}{\overline{\sigma}}
\right)^{d-1}\chi_{0i}\mu_{i0} (1+\alpha_{0i})\left[\frac{\theta_{i}}{\theta_0(\theta_i+\theta_{0})}\right]^{1/2},
\end{equation}
\begin{equation}
\label{a14}
\nu_{c,0i}^*=\frac{2\pi^{(d-1)/2}}{d(d+2)\Gamma\left(\frac{d}{2}\right)}x_i\left(\frac{\sigma_{0i}}{\overline{\sigma}}
\right)^{d-1}\chi_{0i}\mu_{i0} (1+\alpha_{0i})\left(\theta_i+\theta_{0}\right)^{-1/2}\theta_i^{-3/2}\theta_{0}^{1/2}
A_c,
\end{equation}
\begin{equation}
\label{a15}
\nu_{d,0i}^*=\frac{\pi^{(d-1)/2}}{d(d+2)\Gamma\left(\frac{d}{2}\right)}x_i\left(\frac{\sigma_{0i}}{\overline{\sigma}}
\right)^{d-1}\chi_{0i}\mu_{i0} (1+\alpha_{0i})\left(\frac{\theta_0}{\theta_i(\theta_0+\theta_i)}\right)^{3/2}
\left[A_d-(d+2)\frac{\theta_i+\theta_{0}}{\theta_{0}} A_c\right],
\end{equation}
where
\begin{eqnarray}
\label{a16}
A_c&=& (d+2)(2\lambda_{0i}+\theta_i)+\mu_{i0}(\theta_0+\theta_i)\left\{(d+2)(1-\alpha_{0i})
-[(11+d)\alpha_{0i}-5d-7]\lambda_{0i}\theta_0^{-1}\right\}\nonumber\\
& & +3(d+3)\lambda_{0i}^2\theta_0^{-1}+2\mu_{i0}^2\left(2\alpha_{0i}^{2}-\frac{d+3}{2}
\alpha_{0i}+d+1\right)\theta_0^{-1}(\theta_0+\theta_i)^2-
(d+2)\theta_i\theta_0^{-1}(\theta_0+\theta_i),
\eeqa
\beqa
\label{a17}
A_d&=& 2\mu_{i0}^2\theta_0^{-2}(\theta_0+\theta_i)^2
\left(2\alpha_{0i}^{2}-\frac{d+3}{2}\alpha_{0i}+d+1\right)
\left[(d+2)\theta_0+(d+5)\theta_i\right]
\nonumber\\
& &-\mu_{i0}(\theta_0+\theta_i)
\Big[\lambda_{0i}\theta_0^{-2} \big\{(d+2) \theta_0 + (d+5) \theta_i\big\} \big\{(11+d)\alpha_{0i}-5d-7\big\}
\nonumber\\
& &
-\theta_i\theta_0^{-1} \big\{20+d(15-7\alpha_{0i})+d^2(1-\alpha_{0i})-28\alpha_{0i}\big\}
-(d+2)^2(1-\alpha_{0i})\Big]
\nonumber\\
& &+3(d+3)\lambda_{0i}^2\theta_0^{-2}[(d+2)\theta_0+(d+5)\theta_i]+
2\lambda_{0i}\theta_0^{-1}[(d+2)^2\theta_0+(24+11d+d^2)\theta_i]
\nonumber\\
& &+(d+2)\theta_i\theta_0^{-1}
[(d+8)\theta_0+(d+3)\theta_i]-(d+2)(\theta_0+\theta_i)\theta_0^{-2}\theta_i
[(d+2)\theta_0+(d+3)\theta_i].
\eeqa

\end{widetext}

\section{Collision frequency of the intruders}
\label{appC}

In this appendix, we provide the relationship between the dimensionless times $\tau_0(t)$ (defined as the average number of collisions suffered by the intruders up to a time $t$) and $\tau(t)$ [defined by  Eq.\ \eqref{2.10}]. To establish this relation one has to evaluate the (average) collision frequency of the intruders $\nu_0(t)$.
For hard spheres, $\nu_0(t)$ is defined as
\beq
\label{c1}
\nu_0(t)=\sum_{i=1}^s \nu_{0i}(t),
\eeq
where
\beqa
\label{c2}
\nu_{0i}(t)&=&n_0^{-1}\sigma_{0i}^{d-1}\chi_{0i}\int d\mathbf {v}_1 \int d\mathbf{v}_{2}\int d\widehat{\boldsymbol{\sigma}}\,\Theta (\widehat{{\boldsymbol {\sigma}}}\cdot {\bf g}_{12})\nonumber\\
& & \times (\widehat{\boldsymbol {\sigma }}\cdot {\bf g}_{12}) f_i(\mathbf{v}_1;t)f_0(\mathbf{v}_2;t),
\eeqa
where $f_i(\mathbf{v}_1;t)$ and $f_0(\mathbf{v}_2;t)$ are the velocity distribution functions of the particles of the species $i$ and the intruders, respectively. The integrals appearing in Eq.\ \eqref{c2} are evaluated here by considering the Maxwellian approximations to $f_i$ and $f_0$, namely,
\beq
\label{c3}
f_i(\mathbf{v}_1)\to n_i \pi^{-d/2} v_\text{th}^{-d} \theta_i^{d/2}e^{-\theta_ic_1^2},
\eeq
\beq
\label{c3.1}
f_0(\mathbf{v}_2)\to n_0 \pi^{-d/2} \theta_0^{d/2} v_\text{th}^{-d} e^{-\theta_0 c_2^2},
\eeq
where $\mathbf{c}_i=\mathbf{v}_i/v_\text{th}$.  We recall that $v_\text{th}=\sqrt{2T/\overline{m}}$, $\theta_{i}=m_{i}T/(\overline{m}T_{i})$, and $\theta_{0}=m_{0}T/(\overline{m}T_{0})$. Thus, $\nu_{0i}$ can be rewritten as
\beq
\label{c4}
\nu_{0i}(t)=n_i\sigma_{0i}^{d-1}\chi_{0i}  (\theta_i \theta_0)^{d/2} v_\text{th}(t) I_\nu,
\eeq
where we have introduced the dimensionless integral
\beq
\label{c5}
I_\nu=\pi^{-d}\int d\mathbf {c}_1 \int d\mathbf{c}_{2}\int d\widehat{\boldsymbol{\sigma}}\,\Theta (\widehat{{\boldsymbol {\sigma}}}\cdot {\bf g}_{12}^*)(\widehat{\boldsymbol {\sigma }}\cdot {\bf g}_{12}^*)e^{-\theta_ic_1^2-\theta_0 c_2^2}.
\eeq
Here, $\mathbf{g}_{12}^*=\mathbf{g}_{12}/v_\text{th}$. The integral $I_\nu$ can be performed by the change of variables $\mathbf{x}=\mathbf{c}_1-\mathbf{c}_2$, $\mathbf{y}=\theta_i\mathbf{c}_1+\theta_0 \mathbf{c}_2$, with the Jacobian $(\theta_i+\theta_0)^{-d}$. The integral $I_\nu$ gives
\beqa
\label{c6}
I_\nu&=& \pi^{-d}\frac{S_d^2}{\pi^{(d+1)/2}\,\Gamma\left(\frac{d+1}{2}\right)}\,(\theta_i+\theta_0)^{-d} \int_0^\infty dx\, x^d e^{-ax^2}\nonumber\\
& & \times \int_0^\infty dy\, y^{d-1} e^{-b y^2},
\eeqa
where $S_d=2\pi^{d/2}/\Gamma(d/2)$ is the total solid angle in $d$ dimensions, $a\equiv \theta_i\theta_0(\theta_i+\theta_0)^{-1}$,  $b\equiv (\theta_i+\theta_0)^{-1}$ and use has been made of the result \cite{NE98}
\beq
\label{c7}
 \int d\widehat{\boldsymbol{\sigma}}\,\Theta(\widehat{{\boldsymbol {\sigma}}}\cdot {\bf g}_{12}^*)(\widehat{\boldsymbol {\sigma }}\cdot {\bf g}_{12}^*)=\frac{\pi^{(d-1)/2}}{\Gamma\left(\frac{d+1}{2}\right)} \, \mbox{g}_{12}^*.
\eeq
The integral $I_\nu$ gives
\beq
\label{c8}
I_\nu=\frac{\pi^{(d-1)/2}}{\Gamma\left(\frac{d}{2}\right)}(\theta_i\theta_0)^{-\frac{d+1}{2}}\left(\theta_i+\theta_0\right)^{1/2},
\eeq
and hence, $\nu_{0i}$ can be finally written as
\beqa
\label{c9}
\nu_{0i}(t)&=&\frac{\pi^{(d-1)/2}}{\Gamma\left(\frac{d}{2}\right)} n_i\sigma_{0i}^{d-1}\chi_{0i} \left(\frac{\theta_i+\theta_0}{\theta_i\theta_0}\right)^{1/2} v_\text{th}(t)\nonumber\\
&=&\frac{\pi^{(d-1)/2}}{\Gamma\left(\frac{d}{2}\right)} x_i\left(\frac{\sigma_{0i}}{\overline{\sigma}}\right)^{d-1}\chi_{0i} \left(\frac{\theta_i+\theta_0}{\theta_i\theta_0}\right)^{1/2}\nu(t).
\nonumber\\
\eeqa

The (dimensionless) time $\tau_0(t)$ is defined as
\beq
\label{c10}
\tau_0(t)=\int_0^t dt' \nu_0(t')=\sum_{i=1}^s \int_0^t dt' \nu_{0i}(t').
\eeq
Thus, according to Eq.\ \eqref{c9}, the relationship between $\tau_0(t)$ and $\tau(t)$ is
\beq
\label{c11}
\tau_{0}(t)= \int_0^t dt' \, \tau(t')\,  \sum_{i=1}^s A_{0i}(t'),
\eeq
where
\beq
\label{c12}
A_{0i}\equiv \frac{\pi^{(d-1)/2}}{\Gamma\left(\frac{d}{2}\right)} x_i\left(\frac{\sigma_{0i}}{\overline{\sigma}}\right)^{d-1}\chi_{0i} \left(\frac{\theta_i+\theta_0}{\theta_i\theta_0}\right)^{1/2}.
\eeq

%


\end{document}